\documentclass[aps,pra,onecolumn,amsmath,amssymb,footinbib,nopacs,preprint]{revtex4-1}

\newcommand{\Jprl}{Phys. Rev. Lett.}
\newcommand{\Jpr}{Phys. Rev.}
\newcommand{\Jpra}{Phys. Rev. A}
\newcommand{\Jprb}{Phys. Rev. B}

\newcommand{\Jrmp}{Rev. Mod. Phys.}

\newcommand{\Jnature}{Nature}
\newcommand{\Jnatphys}{Nature Phys.}

\newcommand{\Jnjp}{New J. Phys.}
\newcommand{\Jepjd}{Eur. Phys. J. D}

\newcommand{\Jphystoday}{Phys. Today}

\usepackage{graphicx}
\usepackage{amsmath}
\usepackage{amsfonts}
\usepackage{amssymb}
\usepackage[latin1]{inputenc}   
\usepackage{color}
\usepackage{hyperref}
\hypersetup{
colorlinks=true,
citecolor=blue,
linkcolor=red,
urlcolor=black
}

\usepackage{dcolumn}
\usepackage{bm}

\newcommand{\vect}[1]{\mathbf{#1}}

\newcommand{\R}{\mathrm{R}}
\newcommand{\me}{\mathrm{c}}
\newcommand{\perc}{\mathrm{p}}

\newcommand{\floc}{f_\mathrm{loc}}

\newcommand{\lB}{l_{\mathrm{B}}}

\newcommand{\av}[1]{\overline{#1}}

\newcommand{\ti}{t_\mathrm{i}}

\newcommand{\nCOL}{\tilde{n}}
\newcommand{\nCOLi}{\tilde{n}_{\textrm{i}}}

\newcommand{\ProbSpeckle}{\mathcal{P}}

\newcommand{\ri}{\vect{r}_{\textrm{i}}}

\newcommand{\distEri}{\mathcal{D}_{\textrm{i}}}
\newcommand{\distEi}{f_{\textrm{i}}}

\newcommand{\densi}{n_{\textrm{i}}}

\newcommand{\Pdiff}{P}

\begin{document}

\title{Three-dimensional localization of ultracold atoms in an optical disordered potential}

\author{F.~Jendrzejewski$^1$, A.~Bernard$^1$, K.~M\"uller$^1$, P.~Cheinet$^1$, V.~Josse$^1$, M.~Piraud$^1$, L.~Pezz\'e$^1$, L.~Sanchez-Palencia$^1$, A.~Aspect$^1$, and P.~Bouyer$^{1,2}$}
\affiliation{$^1$Laboratoire Charles Fabry UMR 8501,
Institut d'Optique, CNRS, Univ Paris Sud 11,
2 Avenue Augustin Fresnel,
91127 Palaiseau cedex, France \\
$^2$LP2N UMR 5298,
Univ Bordeaux 1, Institut d'Optique and CNRS,
351 cours de la Lib\'eration,
33405 Talence, France.}

\begin{abstract}
We report a study of three-dimensional (3D) localization of ultracold atoms suspended against gravity, and released in a 3D optical disordered potential with short correlation lengths in all directions. We observe density profiles composed of a steady localized part and a diffusive part. Our observations are compatible with the self-consistent theory of Anderson localization, taking into account the specific features of the experiment, and in particular the broad energy distribution of the atoms placed in the disordered potential. The localization we observe cannot be interpreted as trapping of particles with energy below the classical percolation threshold.
\end{abstract}

\date{July 31, 2011}

\maketitle

Anderson localization (AL) was proposed more than 50 years ago~\cite{anderson1958}
to understand how disorder can lead to the total cancellation of conduction in certain materials. It is a purely quantum, one-particle effect, which can be interpreted as due to interference between the various amplitudes associated with the scattering paths of a matter wave propagating among impurities~\cite{lagendijk2009}.
Anderson localization is predicted to strongly depend on the dimension~\cite{abrahams1979}.
In the three-dimensional (3D) case,
a mobility edge is predicted,
which corresponds to an energy threshold separating localized
from extended states.
Determining the precise behavior of the mobility edge remains a challenge for numerical simulations, microscopic theory and experiments~\cite{lagendijk2009}.
The quest
for AL has been pursued not only in condensed matter physics~\cite{lee1985}, but also in wave physics~\cite{tiggelen1999}: for instance with light waves~\cite{wiersma1997,storzer2006,schwartz2007,lahini2008}, microwaves~\cite{dalichaouch1991,chabanov2000} and acoustic waves~\cite{hu2008}. Following theoretical proposals~\cite{damski2003,roth2003,lsp2007,piraud2011a,kuhn2007,skipetrov2008}, recent experiments~\cite{billy2008,roati2008}  have shown that ultracold atoms
in optical disorder
constitute a remarkable system to study 1D localization~\cite{aspect2009,lsp2010,Note1}.
Here, we
report a study of 3D localization of
ultracold
atoms suspended against gravity, and released in a 3D optical disordered potential with short correlation lengths in all directions.
We observe density profiles composed of a steady localized part and a diffusive part.
Our observations are compatible with the self-consistent
theory of AL~\cite{vollhardt1992},
taking into account the specific features of
the experiment, and in particular the
broad energy distribution of the atoms placed in
the disordered potential. The localization we observe cannot be interpreted as trapping of particles with energy below the classical percolation threshold.

\begin{figure*}[t!]
\centering
\includegraphics[width=0.80\textwidth]{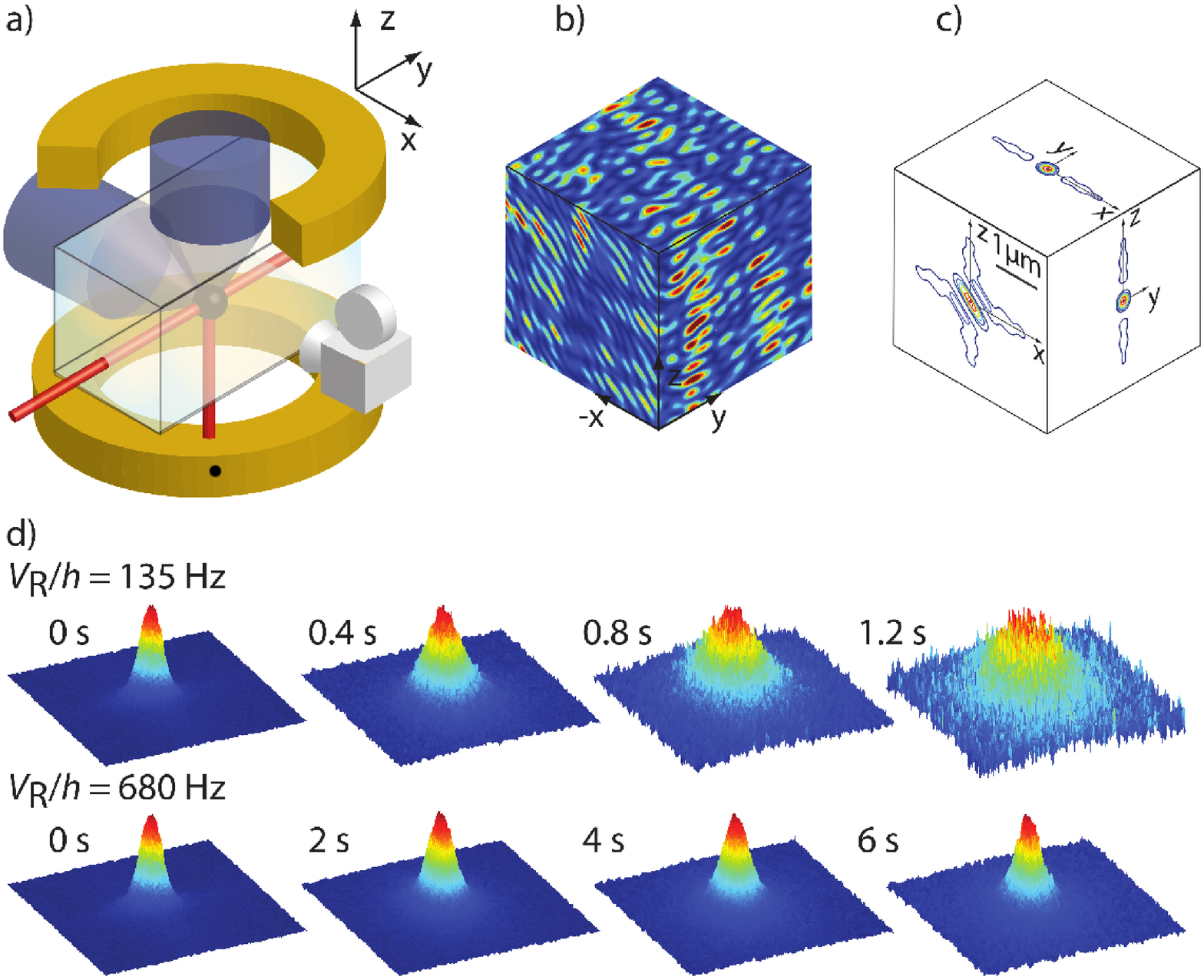}
\caption{\textbf{Experimental set up.}
(a)~A dilute Bose-Einstein Condensate (BEC) of ultra-cold $^{87}\mathrm{Rb}$ atoms, initially trapped by the
red-detuned crossed laser beams, is released and submitted to a repulsive disordered potential
realized by the optical speckle field produced by two crossed, blue-detuned, coherent laser beams along the $x$- and $z$- axes,  passed through diffusive plates.
The  (paramagnetic) atoms are suspended against gravity by a magnetic field gradient (produced by the yellow coils),  and the expansion of the atomic cloud can be observed for times as long as $8$s. (b) Representation (false colors) of the disordered potential in the $x=0$, $y=0$,  and $z=0$ planes.
(c) Plots of the 3D autocorrelation function of the disordered potential in the $x=0$, $y=0$,  and $z=0$ planes (the equal level lines represent levels separated by 14\% of the maximum value).
The correlation radii are
$0.11~\mu$m, $0.27~\mu$m and $0.08~\mu$m, along the main axes (axis $y$ and the two bisecting lines of $x-z$).
(d)~3D expansion of the atomic cloud for two values of the disorder amplitude.
The plots show column density, rescaled to the same maximum, in the $y-z$ plane,
as observed by fluorescence imaging along the $x-$axis at various delays after application of the disorder.
For small disorder ($V_\mathrm{R}/h=135~$Hz), we observe an essentially diffusive expansion.
For larger disorder ($V_\mathrm{R}/h=680~$Hz), a significant fraction of atoms is localized: in the center, the localized atoms emerge in the form of a steady peak, which is a replica of the initial density profile of the BEC.
\label{fig1}}
\end{figure*}

Our scheme (Fig.~\ref{fig1}a) is a generalization of the one that allowed us to demonstrate AL in 1D~\cite{lsp2007,billy2008}. It involves a dilute Bose-Einstein condensate (BEC) with several $10^4$ atoms of $^{87}\mathrm{Rb}$, initially in a shallow quasi-isotropic Gaussian optical trap, released and suddenly submitted to an optical disordered potential generated by a laser speckle~\cite{Note2}. The atoms, in the $\vert F=2, m_F=-2 \rangle$  hyperfine state of the ground electronic state, are suspended by a magnetic gradient that compensates gravity
(the residual component of the magnetic potential is isotropic and repulsive,
of the form $-m\omega^2\vect{r}^2/2$, with $\omega=7$~s$^{-1}$).
In  the experiments reported here, we observe a  uniform loss of atoms, with a decay constant of $\sim 5~\mathrm{s}$,
which we compensate by rescaling all profiles to a fixed total number of atoms.
In the absence of disordered potential, we observe the free ballistic expansion of the BEC,
induced by the initial interaction energy,
as well as the expansion of the thermal wings.
This allows us to determine the maximum velocity $v_\mathrm{max} \sim 0.5$~mm/s in the expanding BEC (corresponding to a chemical potential of the trapped BEC
$\mu_\mathrm{in}=3 m v_\mathrm{max}^2/4$ of the order of $\mu_\mathrm{in}/h \simeq  40$~Hz, where $m$ is the mass of the atom, and $h$ the Planck constant). A Gaussian fit to the velocity distribution in the wings yields a rms velocity of $ \sim 0.3$~mm/s, \textit{i.e.} a  temperature of $T \sim 1~\mathrm{nK}$ ($k_\mathrm{B} T/h \sim 20$~Hz, where $k_\mathrm{B}$ is the Boltzmann constant).
The condensed fraction is $\sim 55\% \pm 5\%$ of the total number of atoms.

The optical disordered potential is switched on, in less than $100\mu$s, at time $\ti=50$~ms after release,
\emph{i.e.}\ in a situation where residual atom-atom interaction energy is small compared to the
disorder amplitude $V_\mathrm{R}$ ($\sim 10$~Hz compared to $100-1000$~Hz.
As shown in ref.~\onlinecite{clement2006}, a well controlled disordered potential can be obtained using the speckle field realized by passing a far detuned laser beam through a diffusive plate~\cite{goodman2007}.
In order to create a 3D disorder  with a small correlation length along any direction of space, we cross two coherent orthogonal speckle fields (see Fig.~\ref{fig1}a).
They have the same polarization (along the $y$-axis), which yields an interference pattern sketched in Fig.~\ref{fig1}b.
The laser has a wavelength of $532$~nm, detuned far below the $^{87}$Rb resonance at $780$~nm, so that the disordered potential is positive or null at any point.
More precisely, the disorder has the single-point probability distribution
$\ProbSpeckle(V) = V_\mathrm{R} \exp({-V/V_\mathrm{R}})$, maximum at $V=0$, and
of average value equal to its standard deviation $V_\mathrm{R}$
(hereafter denoted the "amplitude"), which can be varied up to $V_\R /h = 1.1$~kHz.
Figure~\ref{fig1}c shows cuts of the autocorrelation function of the disorder.
A 3D Gaussian fit of this autocorrelation function yields standard rms radii
of $0.11~\mu$m, $0.27~\mu$m and $0.08~\mu$m, along the main axes (axis $y$ and the two bisecting lines of $x-z$),
with thus a maximum anisotropy factor of 3.
Their geometric average provides the characteristic correlation length
$\sigma_{\R} \simeq 0.13~\mu$m.
The corresponding correlation energy~\cite{kuhn2007}
$E_{\R}=\hbar^2/m \sigma_{\R}^2$  is larger than the disorder amplitude used in the experiment ($E_{\R} /h \simeq  6.5$~kHz, to be compared to $V_\R/h \sim 0.1 - 1.1$~kHz),
so that we work in the quantum disorder regime, which does not support bound states in local minima of the disordered potentials.

\begin{figure}[t!]
\centering
\includegraphics[width=0.7\textwidth]{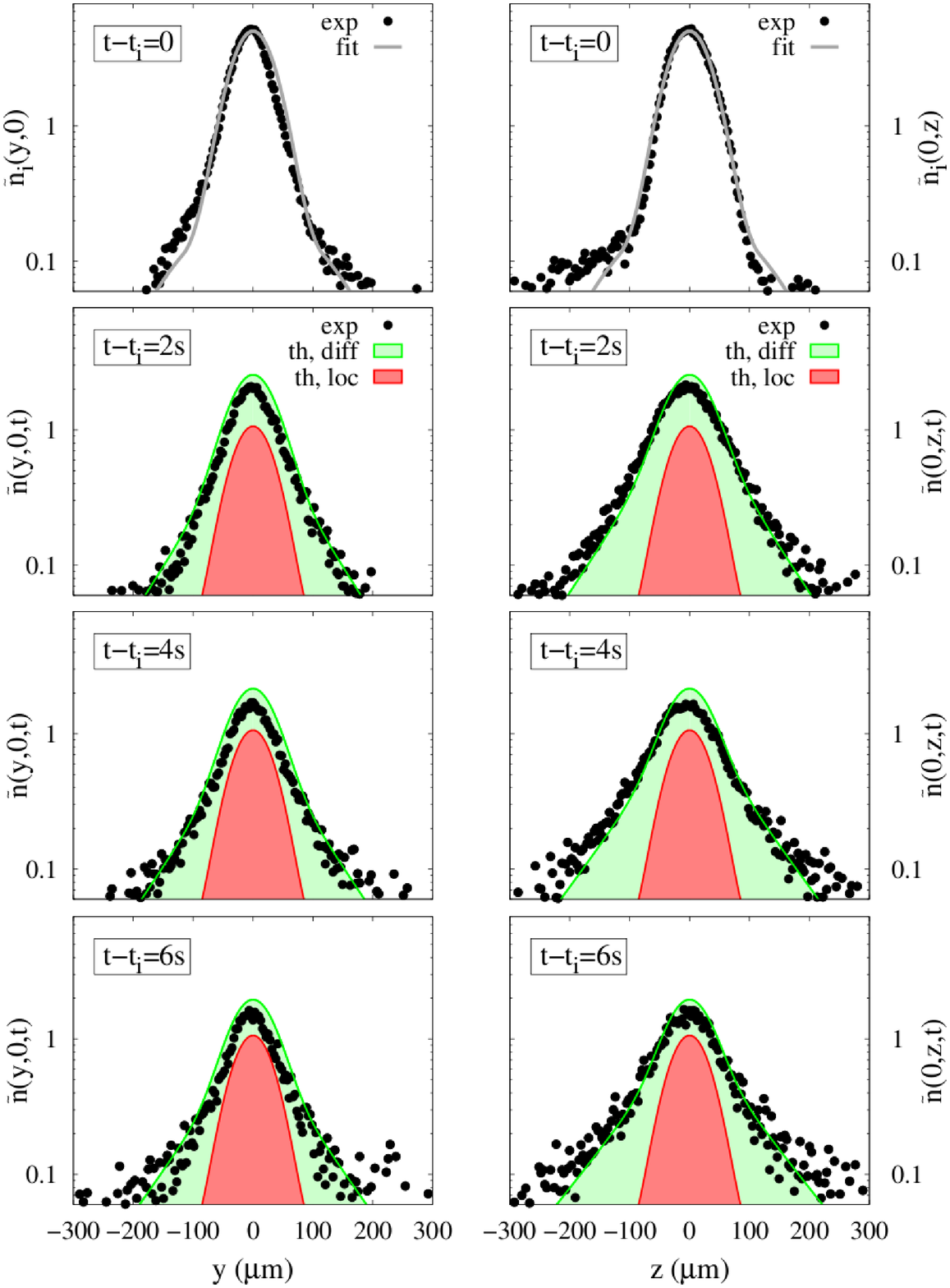}
\caption{\textbf{Density profiles: experiment vs.\ theory.}
The figure shows cuts of the column density profiles along $y$ [$\nCOL(y,0,t)$, left column] and $z$
[$\nCOL(0,z,t)$, right column], imaged at various delays after application of the disorder,
in a situation of significant localization $(V_{\R}/h=680$~Hz).
The black dots are the experimental data.
The solid gray lines in the upper panels are fits to the experimental data at the initial time $\ti$, of the theoretical profile of a BEC in the Thomas-Fermi regime for the optical Gaussian trap, plus the thermal fraction.
This fitted total profile, multiplied by the theoretical localized fraction $\floc$ (see text and purple line in Fig.~\ref{floc}), yields the red profiles, describing the localized contribution.
Adding the theoretically determined diffusive part, we obtain the green profiles,
corresponding to the complete equation~(\ref{eq:DensProf}).
\label{profils}}
\end{figure}

\begin{figure}[t!]
\centering
\includegraphics[width=0.7\textwidth]{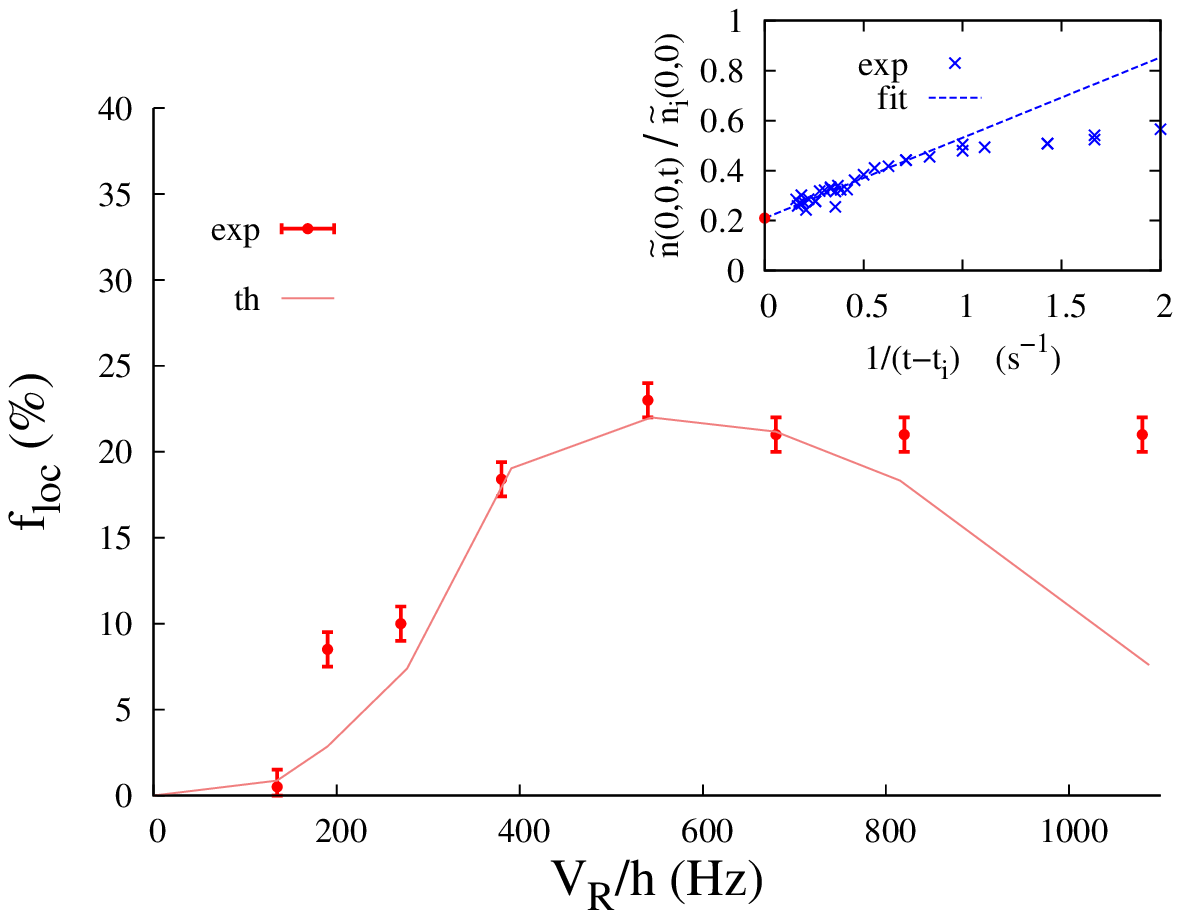}
\caption{\textbf{Localized fraction.}
The inset shows experimental data for the ratio of the column density in the center to its initial value,
$\nCOL (0,0,t)/\nCOLi (0,0)$, versus the inverse time delay after switching on the disorder, $1/(t-\ti)$ (blue points), together with a linear fit at long times (dashed blue line), for $V_{\R}/h=680$Hz. The intercept of this fit with the vertical axis provides the localized fraction, $\floc$ (red point).
The main panel shows the values of $\floc$ obtained by this procedure as a function of the disorder amplitude $V_{\R}$ (red points with errorbars). The solid purple line shows the results of the theoretical calculation, including the heuristic energy shift $+3.35 V_{\R}^2/E_{\R}$ on the energy distribution (see text).
\label{floc}}
\end{figure}

We study how the expansion of the released atomic cloud is affected when we apply the
speckle potential, for several values of the disorder amplitude $V_\mathrm{R}$. Figure~\ref{fig1}d shows the observed column density $\nCOL (y,z,t) = \int d x\ n(x,y,z,t)$ [where $n(x,y,z,t)$ is the atomic density of the expanding cloud], for two different values of the disorder amplitude $V_\mathrm{R}$. For the small value ($V_\mathrm{R}/h= 135$~Hz) the behaviour is essentially diffusive, as can be checked by observing that at long enough times the rms widths $\Delta y$ and $\Delta z$
increase as $(t-t_\mathrm{i})^{1/2}$, and the column density in the center, $\nCOL (y=0,z=0,t)$, decreases
as $(t-t_\mathrm{i})^{-1}$. For the large value of the disorder amplitude ($V_\mathrm{R}/h= 680$~Hz),
the observed density distribution  (see also cuts of the column density along $y$ and $z$ in Fig.~\ref{profils}) is the sum of two contributions with weights of the same order of magnitude: (i)~a steady localized part, which is the replica of the initial profile $\nCOLi (y,z)$, \emph{i.e.}\ the BEC and its thermal wings at $t=t_\mathrm{i}$; (ii)~an evolving diffusive part. More precisely, we have found that we can decompose the observed column density as
\begin{equation}
\nCOL (y,z,t) = \floc \times \nCOLi (y,z) + \nCOL_{\mathrm{D}} (y,z,t)
\label{eq:DensCol}
\end{equation}
where
$\floc$ is the localized fraction,  and $\nCOL_{\mathrm{D}} (y,z,t)$ is the diffusive contribution. This decomposition is consistent with the observation that in the long time limit,  the column density at the center evolves as $\nCOL (0,0,t)/\nCOLi (0,0) = \floc + A/(t-\ti)$, as shown in the inset of Fig.~\ref{floc}. We can then determine $\floc$  as the vertical axis intercept of the plot of $\nCOL (0,0,t)/\nCOLi (0,0)$ versus $1/(t-\ti)$. Figure~\ref{floc} shows the result of that simple analysis of the data,
carried out for different values of $V_\mathrm{R}$. We find a localized fraction starting from zero for vanishing disorder,
then increasing with  $V_\mathrm{R}$, and reaching a nearly saturating value
of $\sim 22\%$ at $V_\mathrm{R}/h \sim 400$~Hz.
As explained in the following, we interpret the localized and diffusive contributions as due to atomic energy components below and above the mobility edge $E_{\me}$, respectively.

In order to compare the experimental results with theory and check whether they are consistent with AL, we use quantum transport theory, taking into account the specific features of our experiment: i)~the broad spatial extension of the atomic gas at the initial time $\ti$;  ii)~its  energy distribution induced by the sudden application of the disordered potential at time $\ti$; iii)~the anisotropy of the 3D speckle potential. We write the spatial density of the atomic gas as~\cite{lsp2007,piraud2011a,skipetrov2008}
\begin{equation}
n (\vect{r},t) = \int d\ri \int dE\ \distEri (\ri,E) \Pdiff (\vect{r} - \ri, t - \ti \vert E) \,,
\label{eq:DensProf}
\end{equation}
where $\distEri (\mathbf{r},E)$ represents the semi-classical joint position-energy density
just after the time $\ti$ when the speckle potential is switched on,
and $\Pdiff (\vect{r}-\ri, t-\ti \vert E)$ is the (anisotropic) probability of quantum diffusion,
\emph{i.e.}\ the probability distribution that a particle of energy $E$, placed in point $\ri$ at time $\ti$,
is found in $\vect{r}$ at $t$. The distribution $\distEri (\vect{r},E)$ depends on both the initial expansion of the atomic gas for $0<t<\ti$, and the disordered potential at $t=\ti^+$.
In the experiment, the sudden application of the disordered potential (in $\sim 100\mu$s)
at time $\ti$ hardly affects the density profile, $\densi (\mathbf{r})$,
but dramatically broadens the energy distribution since the disorder is strong ($V_{\R}^2/E_{\R} \gg \mu_{\mathrm{in}}$).
We thus assume separation of the position and energy variables, \textit{i.e.}\ we write
$\distEri (\mathbf{r},E) = \densi (\mathrm{r}) \times \distEi (E)$.
The initial density profile $\densi (\vect{r})$ is determined from fits to the measured density profile
at time $\ti$ (see Fig.~\ref{profils}).
The energy distribution $\distEi (E)$, averaged over the disorder, is calculated from direct numerical diagonalization of the noninteracting Hamiltonian for various realizations of the disordered potential (see Methods). We find that $\distEi (E)$ is peaked around energy $V_\R$ (the average value of the disordered potential) with a half width at half maximum, ranging from $\sim 10$Hz$\times h$ (for $V_\R / h = 135$Hz) to $\sim 100$Hz$\times h$ (for $V_\R / h = 1088$Hz).

The quantity $\Pdiff (\vect{r}^\prime, t^\prime \vert E)$, whose character changes from localized to extended at the mobility edge $E_\me$, plays the central role in AL. We calculate it, for our 3D speckle potential,  in the framework of the self-consistent approach~\cite{vollhardt1992}:
The incoherent (Boltzmann) diffusion tensor is first evaluated
in the on-shell Born approximation, using quantum transport theory
adapted to disordered potentials with non-isotropic correlation functions~\cite{woelfe1984},
hence generalizing the approach of ref.~\onlinecite{kuhn2007}.
The terms corresponding to the quantum interference between the various diffusing paths are then incorporated in the form of the Cooperon contribution. This provides an equation for the dynamic, quantum corrected diffusion tensor, $D_*(E,\Omega)$, which is finally solved self-consistently in the long time limit ($\Omega \rightarrow 0$).

In the localization regime ($E<E_{\me}$),
$\Pdiff (\vect{r}^\prime, t^\prime \vert E)$
is a static, anisotropic, exponentially localized function
with localization lengths $L_{\mathrm{loc}}^u(E)$ along the transport main axes
(axis $y$ and the two bisecting lines of $x - z$).
We find that $L_{\mathrm{loc}}^u(E)$ is smaller than $5\mu$m, except in a narrow energy window $\Delta E$
close to the mobility edge $E_{\me}$ where it diverges (\emph{e.g.}\ $\Delta E /h \sim 20$Hz for $V_\R / h= 680$Hz).
Since the localization lengths are smaller than the imaging resolution ($5\mu$m) and the initial size of the atomic gas (Thomas-Fermi radius of the BEC $\sim 25~\mu$m) for almost all energy components,
we can safely use $\Pdiff (\vect{r}^\prime \vert E) \simeq \delta (\vect{r}^\prime)$
in equation~(\ref{eq:DensProf}) for $E<E_{\me}$.
This yields a stationary profile, which is simply a replica of the initial profile $\nCOLi (y,z)$
with a weight of $\floc = \int_{-\infty}^{E_{\me}} d E\ \distEi (E)$.
We identify it with the first term in equation~(\ref{eq:DensCol}) and can then compare the measured values of $\floc$ with the result of the theoretical calculation.
Due to the narrow width of the energy distribution (see above), the calculated value of $\floc$ is very sensitive to any approximation in the theoretical calculations (\textit{e.g.}\ on $\distEi(E)$ or $E_{\me}$) and uncertainty on experimental parameters (in particular the amplitude $V_\R$ and the details of the disordered potential)
and we found quantitative discrepancies between experimental values and ab initio calculations of $\floc$.
As shown in Fig.~\ref{floc}, however, a fair agreement is  obtained by introducing heuristically an
energy shift equal to  $+3.35 V_\R^2/E_\R$ (\emph{i.e.}\ $\sim 240$~Hz$\times h$ for $V_\R / h=680$~Hz) on the numerically calculated energy distribution.

\begin{figure}[t!]
\centering
\includegraphics[width=0.7\textwidth]{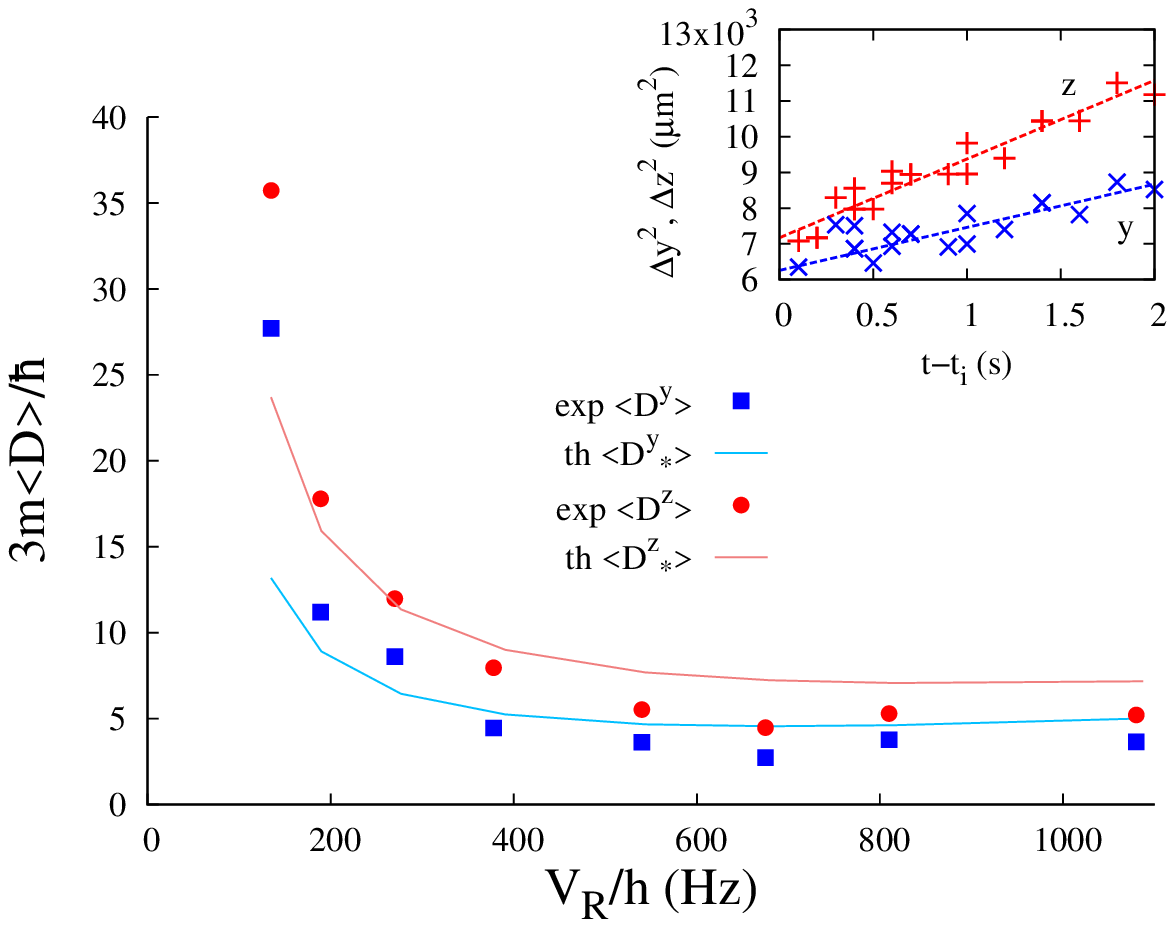}
\caption{\textbf{Diffusive part.}
The main panel shows the average diffusion coefficients, $\langle D^u\rangle$, along the $u=y,z$ axes (blue and red points respectively) versus the disorder amplitude $V_{\R}$. They are obtained from linear fits, $\Delta u^2(t) = 2 \langle D^u \rangle (t-\ti) + \mathrm{const.}$, to plots of the squared rms sizes of the atomic gas, $\Delta y^2$ and $\Delta z^2$, versus the time delay, $t-\ti$
(shown in the inset for $V_{\R}/h=680$Hz).
The corresponding quantum-corrected average diffusion coefficients, $\langle D^u_*\rangle$, obtained from the self-consistent theory of localization are shown as solid cyan ($y$) and purple ($z$) lines.
\label{diffusion}}
\end{figure}

In the diffusion regime ($E>E_{\me}$),
$\Pdiff (\vect{r}^\prime, t^\prime \vert E)$ is an anisotropic Gaussian function
of time-dependent rms widths $\sqrt{2 D_*^u(E) t}$ along the transport main axes.
When incorporated into equation~(\ref{eq:DensProf}), these  components  yield a contribution to the profile that evolves in time. We identify it with the second term in equation~(\ref{eq:DensCol}), whose value at the center vanishes in the long time limit.
In addition, the expansion with time of that diffusive part
allows us to make another comparison between experimental and theoretical results. As shown in the inset of Fig.~\ref{diffusion}, we find that the experimentally measured rms sizes
$\Delta u$, along the $u \in \{y,z\}$ axes,
vary according to
$\Delta u(t)^2=\Delta u(\ti)^2 + 2\langle D^u \rangle (t-\ti)$.
This allows us to determine "average" diffusion coefficients, $\langle D^u \rangle$, which are plotted in  Fig.~\ref{diffusion} versus the disorder amplitude $V_{\R}$.
We can then compare them to the theoretical values for the average quantum corrected diffusion coefficients,
$\langle D_*^u \rangle = \int dE\ \distEi (E) D_*^u (E)$.
Note that while axis $y$ is a transport main axis, axis $z$ is the diagonal between two main axes, so that $D_*^z (E)$ is the half sum of the diffusion coefficients along these two main axes.
As shown in Fig.~\ref{diffusion}, we find a fair agreement between experimental data and theoretical calculations.
In particular, the anisotropy of the diffusion tensor is well reproduced.
Note that the theoretical calculations do not involve any free parameter, apart from the heuristic energy shift ($+3.35V_\R^2/E_\R$) discussed above.

The fair agreement between theory and experiment allows us to interpret the behavior of $\floc$ (Fig.~\ref{floc}) and $\langle D^u \rangle$ (Fig.~\ref{diffusion}),
as resulting from the competition of two effects, when $V_{\R}$ increases.
On the one hand, for each energy component,
the incoherent (Boltzmann) mean free path $\lB(E)$, and thus the diffusion coefficient $D_*^u(E)$, decrease. According to the Ioffe-Regel criterion for localization\cite{Ioffe1960}, $k_E\lB(E) \lesssim 1$
(where $k_E=\sqrt{2mE}/\hbar$ is the typical particle wavevector at energy $E$),
the mobility edge $E_{\me}$ then increases, so that $\floc$ increases.
This effect dominates for weak disorder ($V_\R \lesssim 400$~Hz).
On the other hand, the energy distribution broadens and populates higher energy components.
This latter effect approximately compensates the former one for stronger disorder  ($V_\R \gtrsim 500$ Hz). Hence, the localized fraction reaches a plateau of about $\floc \simeq 22\%$, while  the average diffusion coefficient stops decreasing when it reaches a value  of about $\langle D^{y,z} \rangle \simeq 3-6 \times (\hbar/3m)$.

The experimental results presented here
show a clear evidence that when a large enough disordered potential is applied to an expanding 3D BEC,
a fraction of the atoms (up to $22\%$) gets localized.
It is then natural to ask whether
it corresponds to
3D AL.
To address that question, we rely on various theoretical arguments.
Firstly,
our observations are incompatible with classical localization of particles with an energy below the classical percolation threshold, which is less than $10^{-2}V_{\R}$ for our 3D speckle: the fraction of the atoms with a lower energy is negligible (see Methods).
Actually, in our situation of quantum disorder, the effective percolation threshold may even be lower,
as trapping near potential minima is forbidden.
Secondly, our quantitative measurements of the localized fraction are consistent with a scenario based on the
self-consistent theory of Anderson localization,
applied to the exact experimental situation.
We  obtain a good quantitative agreement, provided we take into account the strong modification of the atoms energy distribution when the disordered potential is applied, as calculated numerically and displaced by a heuristic shift.
A large fraction of the atoms have then an energy above the theoretical value of the mobility edge $E_{\me}$. The calculation is however too sensitive to uncertainties in the experimental parameters, and to approximations in the calculation, to permit a fully quantitative test of the theory.

In order to test precisely
theories of 3D AL, a major experimental improvement would be a method to release, in the disordered potential, a sample  of atoms with a narrow energy distribution, controlled at will, and experimentally measurable. It would then be possible to explore the localization transition,  in particular to measure the exact value of $E_{\me}$, and to determine critical exponents.
A further major change in the experiment will be the control of interactions between the atoms,
since the effect of interactions on AL is an open problem of major interest,
in particular in 3D~\cite{lee1985,lsp2010}.

It will also be important to clarify the status of the energy shift,
which has been introduced heuristically. On one hand, its simple form ($\propto V_\R^2$)
suggests that it may be partially due to some disregarded term at Born order,
for instance the shift of energy states, which is not taken into account in
the on-shell approximation of the self-consistent theory of AL,
but which might be significant~\cite{yedjour2010}.
On the other hand, the above form of the shift may be too simple,
as suggested by the discrepancy with experimental data
obtained for the highest values of $V_\R$ in Fig.~\ref{floc}, and the search of a more elaborated form may lead to a better understanding of the localization phenomenon we have observed.

\section*{Acknowledgements}
We thank S.~Seidel and V.~Volchkov for experimental contributions,
M.~Besbes for assistance on numerical calculations,
and T.~Giamarchi and B.~van~Tiggelen for useful discussions.
This research was supported by
the European Research Council (FP7/2007-2013 Grant Agreement No.\ 256294),
the Agence Nationale de la Recherche (ANR-08-blan-0016-01),
the Minist\`ere de l'Enseignement Sup\'erieur et de la Recherche,
the D\'el\'egation G\'en\'erale de l'Armement,
the Triangle de la Physique
and the Institut Francilien de Recherche sur les Atomes Froids.
We acknowledge the use of the computing facility cluster GMPCS of the 
LUMAT federation (FR LUMAT 2764)

\section*{Methods}

\subsection*{Energy distribution}
Since the initial chemical potential of the BEC ($\mu_\mathrm{in}$) and the thermal energy ($k_\mathrm{B}T$) are smaller than the disorder parameters, the energy distribution can be approximated by $\distEi(E) \simeq A(\vect{k}=0,E)$,
where $A(\vect{k},E) = \langle \vect{k} \vert \av{\delta(E-H)} \vert \vect{k} \rangle$
is the spectral function of the disordered medium, with $H = -\hbar^2\nabla^2/2m + V(\bf{r})$ the noninteracting Hamiltonian associated to a realization of the disordered potential $V(\vect{r})$.
In order to calculate $A(\vect{k}=0,E)$, we decompose the operator $\delta(E-H)$
onto the energy eigenbasis, as obtained by direct numerical diagonalization of the hamiltonian $H$.
The numerical results are obtained in a box of linear length
$\sim 15\lambda$ and of grid step $\sim 0.2\lambda$ ($\lambda = 532$nm is the laser wavelength).
The disorder average is performed over $100$ realizations of $V(\vect{r})$,
with the parameters of the 3D speckle potential used in the experiments.

\subsection*{Percolation.}
The percolation threshold, $E_\perc$, is the energy such that all classical particles of energy $E<E_\perc$ are trapped in finite-size allowed regions.
We have numerically evaluated the percolation threshold of the 3D speckle potential used in the experiment.
Using various values of the grid step, the numerical calculations provide an upper bound for the percolation threshold, $E_\perc \leq 4 (1) \times 10^{-3} V_{\R}$.
Note that, above $E_\perc$, the fraction of classical trapping regions quickly decreases and, according to our numerical calculations, it essentially vanishes for $E \geq 8 (1) \times 10^{-3} V_{\R}$.
Taking into account the energy distribution $\distEi(E)$ calculated numerically,
with or without the heuristic energy shift (see text), we find that the fraction of classically trapped particles is negligible [$\ll 1 \%$, smaller than the numerical resolution of $\distEi(E)$].



%

\end{document}